\documentclass[
reprint,
amsmath,amssymb,
prl,
nofootinbib
]{revtex4-1}
\usepackage{graphicx}
\usepackage{dcolumn}
\usepackage{bm}
\usepackage{color}
\usepackage{hyperref}

\newcommand{\un}[1]{\ensuremath{\,{\rm{#1}}}}

\begin{document}
	
	\title{Stochastic fluctuations of bosonic dark matter}
	
	\author{Gary P. Centers$^{1,2}$,
		John W. Blanchard$^{2}$, 
		Jan Conrad${^3}$,
		Nataniel L. Figueroa$^{1,2}$, 
		Antoine Garcon$^{1,2}$, 
		Alexander V. Gramolin$^{4}$,
		Derek F. Jackson Kimball$^5$, 
		Matthew Lawson$^{2,3}$, 
		Bart Pelssers$^{3}$,
		Joseph A. Smiga$^{1,2}$,
		Alexander O. Sushkov$^{4}$,  
		Arne Wickenbrock$^{1,2}$}
	\author{
		Dmitry~Budker$^{1,2,6,*}$}
	\author{
		Andrei Derevianko$^{7}$}
	\address{$^1$Johannes Gutenberg-Universit{\"a}t, Mainz 55128, Germany}
	\address{$^2$Helmholtz Institute, Mainz 55099, Germany}
	\address{$^3$Department of Physics, Stockholm University, AlbaNova,10691 Stockholm, Sweden}
	\address{$^4$Department of Physics, Boston University, Boston, Massachusetts 02215, USA}
	\address{$^5$Department of Physics, California State University East Bay, Hayward, California 94542-3084,USA}
	\address{$^6$Department of Physics, University of California, Berkeley, CA 94720-7300,USA}
	\address{$^7$Department of Physics, University of Nevada, Reno, Nevada 89557, USA}
	\address{$^*$Corresponding Author, Email: budker@uni-mainz.de}
	
	\date{\today}
	
	\maketitle
	
	\textbf{
		Numerous theories extending beyond the standard model of particle physics predict the existence of bosons \cite{Dimopoulos1996,Arkani-Hamed-2000,Taylor1988,DamourPolyakov1994,Peccei1977b,Peccei1977c,Weinberg1978,Wilczek1978,Lindner2015,Irastorza2018} that could constitute the dark matter (DM) permeating the universe. 
		In the standard halo model (SHM) of galactic dark matter the velocity distribution of the bosonic DM field defines a characteristic coherence time $\tau_c$. 
		Until recently, laboratory experiments searching for bosonic DM fields have been in the regime where the measurement time $T$ significantly exceeds $\tau_c$~\cite{DePanfilis1987,Wuensch1989,Hagmann1990,Asztalos2010,Graham2013a,Budker2014,Brubaker2017,Caldwell2017,Miller2017,Chung2015,Choi2017a,McAllister2017,Alesini2017,Stadnik2015,grote2019novel}, so null results have been interpreted as constraints on the coupling of bosonic DM to standard model particles with a bosonic DM field amplitude $\Phi_0$ fixed by the average local DM density. 
		However, motivated by new theoretical developments \cite{Marsh2016,Marsh2014,Hu2000,Hui2017,Arvanitaki2010,ArvHuaTil15}, a number of recent searches \cite{Abel2017,garcon2019constraints,wu2019search,terrano2019constraints,VanTilburg2015,Hees2016,Wciso2018} probe the regime where $T\ll\tau_c$. 
		Here we show that experiments operating in this regime do not sample the full distribution of bosonic DM field amplitudes and therefore it is incorrect to assume a fixed value of $\Phi_0$ when inferring constraints on the coupling strength of bosonic DM to standard model particles. 
		Instead, in order to interpret laboratory measurements (even in the event of a discovery), it is necessary to account for the stochastic nature of such a virialized ultralight field (VULF) \cite{GeraciDerevianko2016-DM.AI,Derevianko2018}. The constraints inferred from several previous null experiments searching for ultralight bosonic DM were overestimated by factors ranging from 3 to 10 depending on experimental details, model assumptions, and choice of inference framework.
	}

	%
	%

	It has been nearly ninety years since strong evidence of the missing mass we label today as dark matter was revealed \cite{Zwicky1933}, and its composition remains one of the most important unanswered questions in physics.
	There have been many DM candidates proposed and a broad class of them, including scalar (dilatons and moduli \cite{Dimopoulos1996,Arkani-Hamed-2000,Taylor1988,DamourPolyakov1994}) and pseudoscalar particles (axions and axion-like particles \cite{Peccei1977b,Peccei1977c,Weinberg1978,Wilczek1978,Lindner2015,Irastorza2018}), can be treated as an ensemble of identical bosons, with statistical properties of the corresponding fields described by the SHM~\cite{Kuhlen2014,Freese2013a}. 
	In this work, our model of the resulting bosonic field assumes that the local DM is virialized and neglects non-virialized streams of DM~\cite{Diemand2008}, Bose-Einstein condensate formation~\cite{Sikivie2009,Davidson2015,Berges2015}, and possible small-scale structure such as miniclusters~\cite{JacksonKimball2018,Khlopov1985}. To date it is typical to ignore such DM structure when calculating experimental constraints, and we demonstrate the general weakening of inferred constraints due to the statistical properties of the VULF within this isotropic SHM DM model. We note that including sub-halo structure~\cite{Knirck2018,Foster2018}, the formation of which is demonstrated in Refs.~\cite{Chan2018,Lin2018,Veltmaat2018}, can also affect experimental constraints.
	
	
	During the formation of the Milky Way the DM constituents relax into the gravitational potential and obtain, in the galactic reference frame, a velocity distribution with a characteristic dispersion (virial) velocity $v_\mathrm{vir} \approx10^{-3} c$ and a cut-off determined by the galactic escape velocity. 
	Following Refs.~\cite{GeraciDerevianko2016-DM.AI,Derevianko2018} we refer to such virialized ultralight fields, $\phi({t,\boldsymbol{r}})$, as VULFs, emphasizing their SHM-governed stochastic nature.
	Neglecting motion of the DM, the field oscillates at the Compton frequency $ f_{c} = m_{\phi} c^{2}h^{-1}$. However, there is broadening due to the SHM velocity distribution according to the dispersion relation for massive nonrelativistic bosons:  $f_{\phi} = f_{c}  + m_{\phi} v^{2}(2h)^{-1}$. 
	The field modes of different frequency and random phase interfere with one another resulting in a net field exhibiting stochastic behavior.
	The dephasing of the net field can be characterized by the coherence time~\footnote{We note that there is some ambiguity in the definition of the coherence time, up to a factor of 2$\pi$, and adopt that which was used in the majority of the literature. See the discussion in the Supplementary Material.} $\tau_{c}\equiv\left(f_{c} v_{\mathrm{vir}}^{2}/c^{2} \right)^{-1}$~\cite{Schive2014}.
	
	While the stochastic properties of similar fields have been studied before, for example in the contexts of statistical radiophysics, the cosmic microwave background, and stochastic gravitational fields~\cite{RomanoCornish2017}, the statistical properties of VULFs have only been explored recently. 
	The 2-point correlation function, $\langle \phi(t, \mathbf{r}) \phi(t', \mathbf{r}') \rangle$, and corresponding frequency-space DM ``lineshape'' (power spectral density, PSD) were derived in Ref.~\cite{Derevianko2018}, and rederived in the axion context by the authors of Ref.~\cite{Foster2018}. While Refs.~\cite{Derevianko2018,Foster2018} explicitly discuss data-analysis implications in the regime of the total observation time $T$ being much larger than the coherence time, $T\gg\tau_c$, detailed investigation of the regime $T\ll\tau_c$ has been lacking (although we note that Ref.~\cite{Foster2018} includes a brief discussion of the change in sensitivity~\footnote{
				The authors discuss the change of sensitivity due to coherent averaging of the signal in the $T\ll\tau_c$ regime, $T^{1/4}\rightarrow T^{1/2}$, in their Appendix E. There is no mention of how the velocity and amplitude distributions would impact the derived limits.}
		in this regime).

	Here we focus on this regime, $T\ll\tau_c$, characteristic of experiments searching for ultralight (pseudo)scalars with masses $\lesssim 10^{-13}$~eV~\cite{Abel2017,garcon2019constraints,wu2019search,terrano2019constraints,VanTilburg2015,Hees2016,Wciso2018} that have field coherence times $\gtrsim1~\textrm{day}$. This mass range is of significant interest as the lower limit on the mass of an ultralight particle extends to $10^{-22} \un{eV}$ and can be further extended if it does not make up all of the DM \cite{Marsh2016}. Additionally, there has been recent theoretical motivation for ``fuzzy dark matter'' in the $10^{-22}-10^{-21} \un{eV}$ 
	range~\cite{Marsh2016,Marsh2014,Hu2000,Hui2017} and the so-called string ``axiverse'' extends to $10^{-33} \un{eV}$ \cite{Arvanitaki2010}. Similar arguments also apply to dilatons and moduli~\cite{ArvHuaTil15}. 
	
	\begin{figure}
		\includegraphics[width=\linewidth]{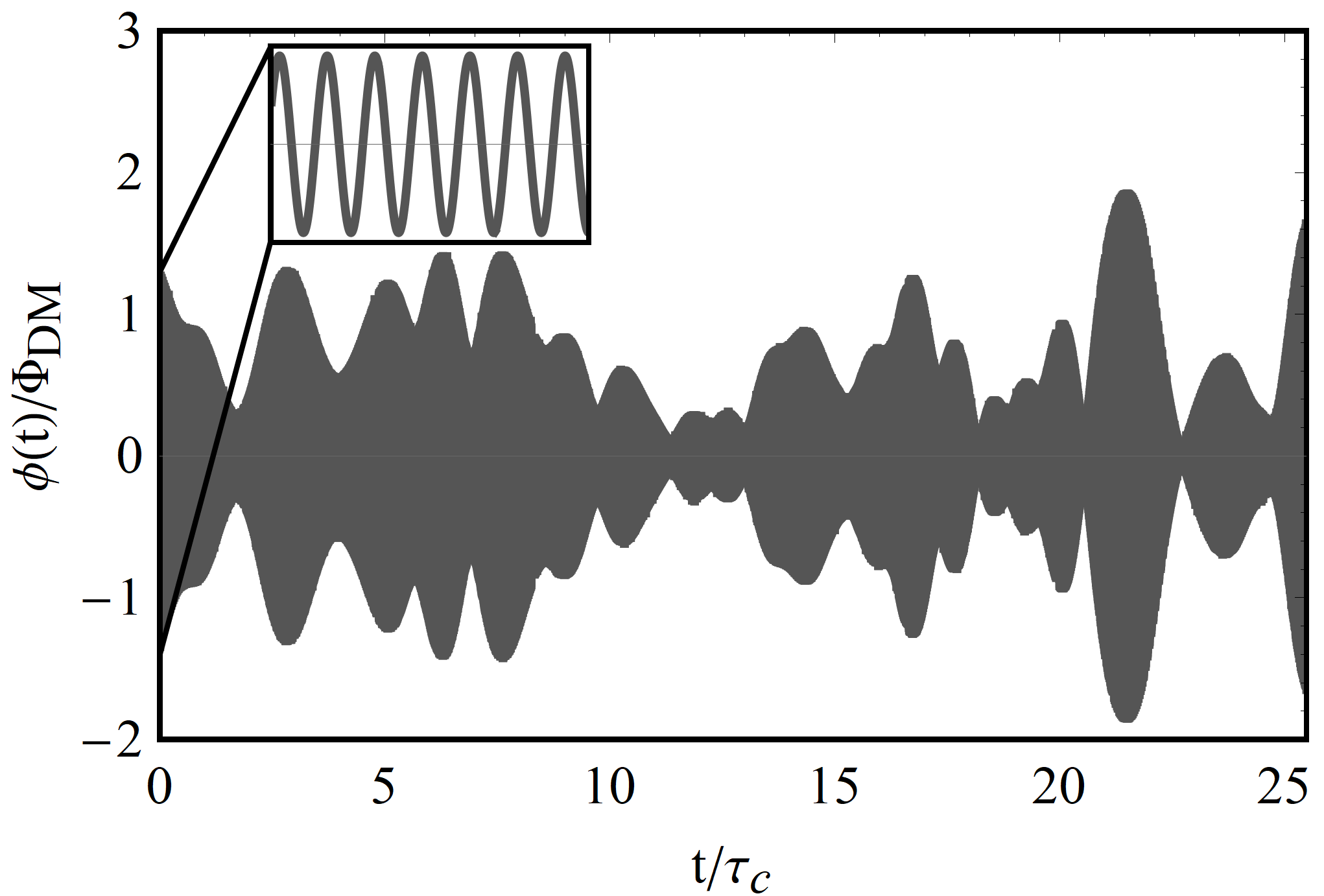}
		\caption{Simulated VULF based on the approach in Ref.~\cite{Derevianko2018} with field value $\phi(t)$ and time normalized by $\Phi_{{\rm{DM}}}$ and coherence time $\tau_c$ respectively. The inset plot displays the high-resolution coherent oscillation starting at $t=0$.}
		\label{Fig:VULFsimulation}
	\end{figure}
	
	Figure 1 shows a simulated VULF field, illustrating the amplitude modulation present over several coherence times.
	At short time scales ($\ll \tau_{c}$) the field coherently oscillates at the Compton frequency, see the inset of Fig. \ref{Fig:VULFsimulation}, where the amplitude $\Phi_0$ is fixed at a single value sampled from its distribution. An unlucky experimentalist could even have near-zero field amplitudes during the course of their measurement.
	
	On these short time scales the DM signal $s(t)$ exhibits a harmonic signature, 
	\begin{equation}
	s(t) =\gamma \xi\phi(t) \approx \gamma\xi \Phi_{0} \cos( 2 \pi f_{\phi} t + \theta) \,,
	\label{Eq:signal}
	\end{equation} 
	where $\gamma$ is the coupling strength to a standard-model field and $\theta$ is an unknown phase. Details of the particular experiment are accounted for by the factor $\xi$.
	In this regime the amplitude $\Phi_{0}$ is unknown and yields a time-averaged energy density $\langle \phi(t)^{2} \rangle_{T \ll \tau_{c}} = \Phi_{0}^{2}/2$. However, for times much longer than $\tau_c$ the energy density approaches the ensemble average determined by $\langle \Phi_{0}^{2} \rangle = \Phi_\mathrm{DM}^{2}$. This field oscillation amplitude is estimated by assuming that the average energy density in the bosonic field is equal to the local DM energy density $\rho_\mathrm{DM}\approx 0.4 \un{GeV/cm^{3}}$, and thus $ \Phi_\mathrm{DM} =\hbar(m_{\phi}c)^{-1}\sqrt{2\rho_\mathrm{DM}}$.
	
	The oscillation amplitude sampled at a particular time for a duration $\ll \tau_c$ is not simply $\Phi_\mathrm{DM} $, but rather a random variable whose sampling probability is described by a distribution characterizing the stochastic nature of the VULF. Until recently, most experimental searches have been in the $m_\phi\gg10^{-13} \un{eV}$ regime with short coherence times $\tau_c \ll 1\un{day}$. However, for smaller boson masses it becomes impractical to sample over many coherence times: for example,  $\tau_c \gtrsim 1 \un{year}$ for $m_\phi \lesssim 10^{-16} \un{eV}$. 
	Assuming the value $\Phi_{0}=\Phi_{{\rm{DM}}}$ neglects the stochastic nature of the bosonic dark matter field \cite{Abel2017,garcon2019constraints,wu2019search,terrano2019constraints,VanTilburg2015,Hees2016,Wciso2018}.
	
	The net field $\phi(t)$ is a sum of different field modes with random phases. The oscillation amplitude, $\Phi_0$, results from the interference of these randomly phased oscillating fields. This can be visualized as arising from a random walk in the complex plane, described by a Rayleigh distribution~\cite{Foster2018}
	\begin{equation}
	p({\Phi _0}) =
	{\frac{{2{\Phi _0}}}{{\Phi _{{\rm{DM}}}^2}}\exp \left( { - \frac{{\Phi _0^2}}{{\Phi _{{\rm{DM}}}^2}}} \right) \,,}
	\label{Eq:Rayleigh}
	\end{equation}
	analogous to that of chaotic (thermal) light~\cite{Loudon83_book}.
	This distribution implies that $\approx$~63\% of all amplitude realizations will be below the r.m.s.~value $\Phi_\mathrm{DM}$. Equation~\eqref{Eq:Rayleigh}~\cite{Foster2018} is typically represented in its exponential form~\cite{Knirck2018} (see Supplementary Material), and is well sampled in the $T\gg\tau_c$ regime. However, this stochastic behavior should not be ignored in the opposite limit.
	
	We refer to the conventional approach assuming $\Phi_{0}=\Phi_{{\rm{DM}}}$ as \textit{deterministic} and approaches that account for the VULF amplitude fluctuations as \textit{stochastic}. To compare these two approaches we choose a Bayesian framework and calculate the numerical factor affecting coupling constraints, allowing us to illustrate the effect on exclusion plots of previous deterministic constraints~\cite{Abel2017,garcon2019constraints,wu2019search,terrano2019constraints,VanTilburg2015,Hees2016,Wciso2018}. It is important to emphasize that different frameworks to interpret experimental data than presented here can change the magnitude of this numerical factor \cite{Protassov_2002,Cowan2011,Conrad2015,Tanabashi2018}, see Supplementary Material for a detailed discussion. In any case, accounting for this stochastic nature will generically relax existing constraints as we show below.
	
	{\em Establishing constraints on coupling strength  --- }
	We follow the Bayesian framework~\cite{Gre10book} (see application to VULFs in Ref.~\cite{Derevianko2018}) to determine constraints on the coupling-strength parameter $\gamma$. 
	Bayesian inference requires prior information on the parameter of interest to derive its respective posterior probability distribution, in contrast to purely likelihood-based inference methods. 
		The central quantity of interest in our case is the posterior distribution for possible values of $\gamma$, derived from Bayes theorem,
		\begin{align}
		p(\gamma| D, f_{\phi}, \xi) = \mathcal{C} &\int p(\gamma, \Phi_{0} )  \mathcal{L}( D| \gamma, \Phi_{0} , f_{\phi}, \xi) d\Phi_{0}\, . \label{Eq:PosteriorDef}
		\end{align}
		The left-hand side of the equation is the posterior distribution for $\gamma$, where $D$ represents the data, and the Compton frequency $f_{\phi}$ is a model parameter. $\mathcal{C}$ is the normalization constant, and
		the likelihood $\mathcal{L}(\cdots)$ is the probability of obtaining the data $D$ given that the model and prior information, such as those provided by the SHM, are true.
		The integral on the right-hand side accounts for (marginalizes over) the unknown VULF amplitude $\Phi_{0}$, which we assume follows the Rayleigh distribution described by Eq.~(\ref{Eq:Rayleigh}). For the choice of prior $p(\gamma,\Phi_0)$ we use what is known as an objective prior~\cite{Kass1996}: the Berger-Bernardo reference prior~\footnote{This approach is equivalent to starting with the marginal likelihood $\int d\Phi_0 p(\Phi_0)\mathcal{L(\cdots)}$ and using Jefferey's prior to calculate the posterior~\cite{Bernardo1979}. See details in the Supplementary Material.}~\cite{Berger1992a}. 
		Results from Bayesian inference are sensitive to the choice of prior~\cite{Berger1992a}, and we find better agreement with frequentist based approaches when using an objective prior rather than a uniform prior $p(\gamma)=1$ (see Supplementary Material).
	
	It is important to note that experiments searching for couplings of VULFs to fermion spins (axion ``wind'' searches) are sensitive not only to the amplitude of the bosonic filed but also to the relative velocity between the laboratory and the VULF, which stochastically varies on a time scale $\approx\tau_c$~\cite{Graham2013a}. The signal due to the axion wind is proportional to the projection of this stochastically varying velocity onto the sensitive axis of the experiment. Accounting for the stochastic nature of the relative velocity increases the uncertainty of the derived coupling strength for a given measurement. 
	Axion-wind experiments can also utilize the daily modulation of this projection, due to rotation of the Earth, to search for signals with an oscillation period much longer than the measurement time $T\ll1/f_\phi$. The unknown initial phase $\theta$ of the VULF sets the amplitude of this daily oscillation and also needs to be marginalized over. We describe how we account for stochastic variations of velocity and daily modulation in the relevant experiments~\cite{Abel2017,garcon2019constraints,wu2019search,terrano2019constraints} in the Supplementary Material and focus solely on the stochastic variations of the amplitude, $\Phi_0$, here.
	
	Using the posterior distribution, $p(\gamma| D, f_\phi, \xi)$, one can set constraints on the coupling strength $\gamma$. Such a constraint at the commonly employed 95\% confidence level (CL), $\gamma_{95\%}$, is given by 
	\begin{equation}
	\int_{0}^{\gamma_{95\%}}  p(\gamma| D, f_\phi, \xi)  d\gamma = 0.95 \, .
	\label{Eq:95CLdef}
	\end{equation}
	
\begin{figure}
		\includegraphics[width=\linewidth]{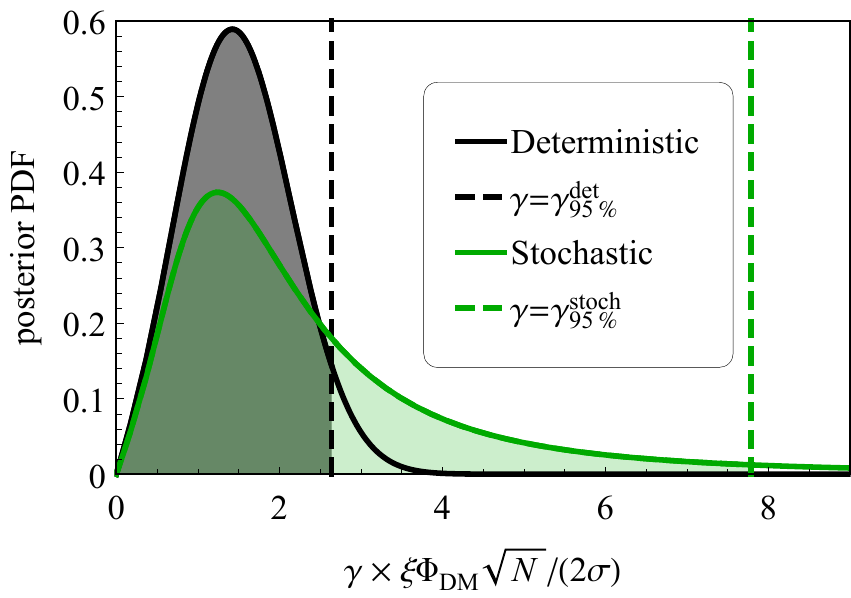}
		\caption{Posterior distributions for the coupling strength $\gamma$ in the deterministic and stochastic treatments, Eqs.~\eqref{Eq:PostDet} and~\eqref{Eq:PostStoch} respectively. Due to the fat-tailed shape of the stochastic posterior one can clearly see the 95\% limit is larger with $\gamma_{95\%}^\mathrm{stoch}/ \gamma_{95\%}^\mathrm{det}\approx3.0$. The assumed value of the data is at the 95\% detection threshold $A^{dt}=\sqrt{-\ln{(1-0.95)}}$ (see text).}
		\label{Fig:posteriors}
	\end{figure}

	The posteriors in both the deterministic and stochastic treatments are derived in the Supplementary Material. 
		In short, the two posteriors differ due to the marginalization over $\Phi_0$ for the stochastic case, see the integral of Eq.~\eqref{Eq:PosteriorDef}. Assuming white noise of variance $\sigma^{2}$ and that the data are in terms of excess amplitude $A$ (observed Fourier amplitude divided by expected noise, an analog to the excess power statistic) we can derive the posterior for excess signal amplitude $A_s$. The posteriors are
		\begin{align}
	&	p_\mathrm{det}(A_s\vert A)  \propto  p(A_s)2A\exp \left( -A^2-A_s^{2}  \right) 
		I_{0} \left(2 A A_s \right) \, ,
		\label{Eq:PostDet} \\
	&	p_\mathrm{stoch}(A_s \vert A) \propto 
		p(A_s)\frac{2A }{(1 + A_s^{2}) } \exp\left( -\frac{A^2}{1 + A_s^{2}    }  \right) \, .
		\label{Eq:PostStoch} 
		\end{align}
		Here $A_s \equiv \gamma  \times  \xi \Phi_\mathrm{DM}\sqrt{N}/(2\sigma)$, $I_{0}(x)$ is the modified Bessel function of the first kind, and $p(A_s)$ is effectively the prior on $\gamma$. In Fig.~\ref{Fig:posteriors} we plot the normalized posteriors assuming $A$ at the 95\% detection threshold $A^{dt}=\sqrt{-\ln{(1-0.95)}}$ and using Berger-Bernardo reference priors for $p(A_s)$; we compare other choices of prior in the Supplementary Material.
		The derivation relies on the discrete Fourier transform for a uniform sampling grid of $N$ points and the assumptions of the uniform grid and white noise can be relaxed~\cite{Derevianko2018}.
	
	Examination of Eqs.~(\ref{Eq:PostDet}),~(\ref{Eq:PostStoch}) and Fig.~\ref{Fig:posteriors} reveals that the fat-tailed stochastic posterior is much broader than the Gaussian-like deterministic posterior. 
	It is clear that for the stochastic posterior, the integration must extend considerably further into the tail, leading to larger values of $\gamma_{95\%}$ and thereby to weaker constraints, $\gamma_{95\%}^\mathrm{stoch} > \gamma_{95\%}^\mathrm{det}$. 
	Explicit evaluation of Eq.~(\ref{Eq:95CLdef}) with the derived posteriors results in a relation between the constraints
	\begin{equation}
	\gamma_{95\%}^\mathrm{stoch}  \approx 3.0  \, \gamma_{95\%}^\mathrm{det} \,, 
	\label{Eq:95CLStocVsDet}
	\end{equation}
	where the numerical value of the correction factor depends on CL and assumed value of $A$ (the factor increases for higher CL and decreases for smaller $A$).
	
	This correction factor becomes $\approx10$ when derived using a uniform prior, as discussed in the Supplementary Material. However, the result obtained with the uniform prior is not invariant under a change of variables (e.g. from excess amplitude to power). Additionally, using the objective prior yields better agreement with frequentist-based results of a factor $\approx2.7$. For the pseudoscalar coupling, the additional stochastic parameters (field velocity and phase) increase this factor up to $\approx8.4$ as shown in the Supplementary Material.

\begin{figure}[h!]
		\includegraphics[width=\linewidth]{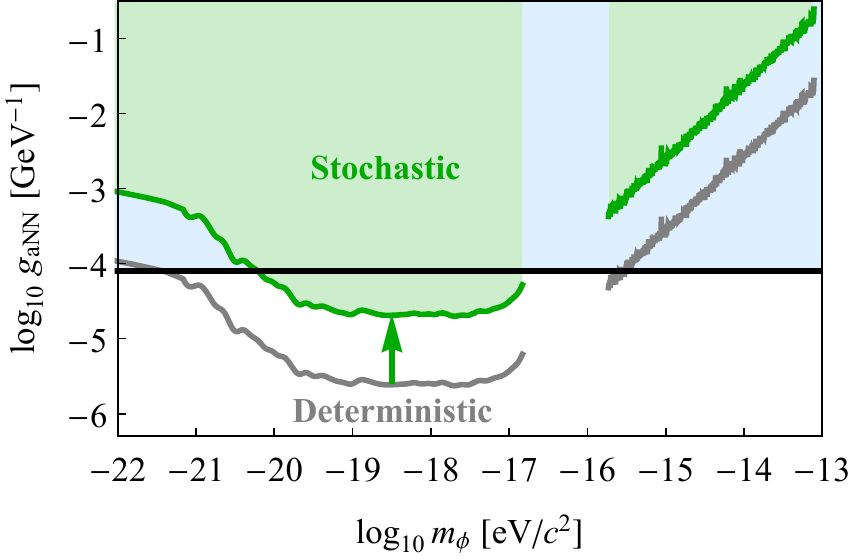}
		\caption{The modified constraint, green line, based on the stochastic approach compared to previous laboratory constraints, gray line, based on the deterministic approach for the axion-nucleon coupling strength $g_{aNN}$~\cite{Abel2017,wu2019search,garcon2019constraints,Vasilakis2009}. The black line represents a constraint from a new-force search using a K-$^3$He comagnetometer~\cite{Vasilakis2009}, unaffected by the local galactic properties of DM.}
		\label{Fig:axionexclusioncombined}
	\end{figure}

	Ultralight DM candidates are theoretically well motivated and an increasing number of experiments are searching for them. 
	Most of the experiments with published constraints thus far are haloscopes, sensitive to the local galactic DM and affected by Eq.~\eqref{Eq:95CLStocVsDet}. 
	However, experiments that measure axions generated from a source, helioscopes or new-force searches, for example, do not fall under the assumptions made here. 
	We illustrate how the existing constraints have been affected in Fig.~\ref{Fig:axionexclusioncombined} and provide more detailed exclusion plots for the axion-nucleon coupling $g_{aNN}$~\cite{Abel2017,wu2019search,garcon2019constraints,Vasilakis2009} and for dilaton couplings~\cite{Hees2016,VanTilburg2015,Wciso2018} in the Supplementary Material.
	
	Figure 3 shows that a few previously published constraints for the axion-nucleon coupling~\cite{Abel2017,wu2019search,garcon2019constraints} no longer constrain new parameter space with respect to the new force constraint at $\approx10^{-4}$ GeV$^{-1}$~\cite{Vasilakis2009}.
	
	
	{\em Conclusion -- }
	To interpret the results of an experiment searching for bosonic DM in the regime of measurement times smaller than the coherence time, stochastic properties of the net field must be taken into account.
	An accurate description accounts for the Rayleigh-distributed amplitude $\Phi_0$, where the variation is induced by the random phases of individual virialized fields. 
	Accounting for this stochastic nature yields a correction factor of $\approx2.7-10$, relaxing existing experimental bosonic DM constraints in this regime.
	In the event of a bosonic DM discovery, the stochastic properties of the field would result in increased uncertainty in the determination of coupling strength or local average energy density in this regime.
	
	It is important to note that observational knowledge of the local distribution of DM can constrain stochastic behavior of the amplitude (energy density). The smallest features observed so far are on the order of $\approx10$ pc~\cite{Iocco2015} (corresponding to a $m_\phi\approx10^{-21}$ eV coherence length), however the analysis in Ref.~\cite{Iocco2015} performs radial averages which would remove the stochastic variation discussed in this paper.
	
%

	{\em Data Availability -- }
	All conclusions made in this paper can be reproduced using the information presented in the manuscript and/or Supplementary Material. Additional information is available upon reasonable request to the corresponding author. For access to the experimental data presented here please contact the corresponding authors of the respective papers.
	
	{\em Acknowledgments -- }
	We thank Eric Adelberger and William A. Terrano for pointing out the need to account for the unknown phase in the CASPEr-ZULF Comagnetometer analysis. We thank Kent Irwin, Marina Gil Sendra, and Martin Engler for helpful discussions and suggestions.
	We thank M. Zawada, N. A. Leefer, and A. Hees for providing raw data for the published deterministic constraints. We also thank Jelle Aalbers for helpful discussions and expert advice on the blueice inference framework.
	Jan Conrad appreciates the support by the Knut and Alice Wallenberg Foundation.
	This project has received funding from the European Research Council (ERC) under the European Unions Horizon 2020 research and innovation programme (grant agreement No 695405). 
	We acknowledge the partial support of the U.S. National Science Foundation, the Simons and Heising-Simons Foundations, and the DFG Reinhart Koselleck project. 
	\bibliography{Amplitude_Fluctuations_in_Bosonic_Dark_Matter,library-apd}
\end{document}


\title{Supplementary material}

\maketitle

\section{Calculating the exclusion plot revision factor}

Ultralight dark matter fields are composed of a large number of individual modes with randomized phases. Thus, as noted in the main text, due to the central-limit theorem, such virialized ultra light fields (VULFs) belong to a broad class of Gaussian random fields that are encountered in many subfields of physics. The statistical properties of such fields are fully described by the two-point correlation functions in the time domain or power-spectral densities in frequency space. A VULF-specific discussion is presented in Ref.~\cite{Derevianko2018}.

Here we consider ultralight dark matter searches in the regime where the observation time $T$ is shorter than the coherence time $\tau_{c}$ of dark matter field.  Because the VULF oscillations are nearly coherent on these short time scales, the VULF-generated signal $s(t)$ in a detector can be described by an oscillatory function
\begin{equation}
s(t) = \gamma \xi \phi(t) = \gamma\, \xi\, \Phi_{0} \, \cos( 2\pi  f_{\phi} t +\theta) \label{Eq:Signal} \, .
\end{equation}
Here $f_{\phi}$ is the field frequency (specific to a particular realization of the VULF as the Compton frequency, $f_c$, is Doppler shifted by the relative motion with respect to the detector, $f_\phi = f_c +m_\phi v^2/(2h)$), $\theta$ is an unknown yet fixed phase, $\xi$ is a constant determined by the details of the experiment, and $\gamma$ is a coupling strength that is, assuming a null result,  to be constrained by the measurement. We fix  the amplitude of the DM field to be positive, $\Phi_{0} \ge 0$ and  the phase $\theta$ of the field to lie in the range $[0 -2\pi)$. The coupling strength $\gamma$ can have an arbitrary sign.

In the conventional approach, the DM field amplitude $\Phi_{0}$  is fixed to its root-mean-square value $\Phi_\mathrm{DM}$,  defined in the main main body of the paper. 
We  refer to such an approach as {\em deterministic}. The key issue is that  for a given experimental run $T \ll \tau_{c}$, the value of the VULF amplitude $\Phi_{0}$  can substantially differ from  $\Phi_\mathrm{DM}$; the VULF amplitude randomly fluctuates. 
This  is apparent from the simulation presented in Fig.~1 of the main text.  
Thus $\Phi_{0}$ has to be treated as a nuisance parameter in the analysis.  We refer to this treatment as {\em stochastic}. This has important implications for establishing constraints on $\gamma$. 
 
 The probability densities for the DM field amplitude  in the two approaches can be summarized as 
 \begin{equation}
 p({\Phi _0}|\Phi_\mathrm{DM}) = \left\{ {\begin{array}{*{20}{c}}
{\delta ({\Phi _0} - {\Phi _{{\rm{DM}}}})\,,}&{{\rm{deterministic}}}\\
{\frac{{2{\Phi _0}}}{{\Phi _{{\rm{DM}}}^2}}\exp \left( { - \frac{{\Phi _0^2}}{{\Phi _{{\rm{DM}}}^2}}} \right) \, ,}&{{\rm{stochastic}}}
\end{array}} .\right. \label{Eq:AmplDistros}
 \end{equation} 
 
 The stochastic probability distribution for the field amplitude can be recast into the probability distribution for the local energy density $\rho = \Phi_{0}^{2} f_{c}^{2}/2$ (where $f_c$ is the Compton frequency, not the Doppler shifted $f_\phi$) of the dark matter field, 
 \begin{equation}
 p(\rho|\rho_\mathrm{DM}) d\rho = \frac{1}{\rho_\mathrm{DM} } e^{- \rho/\rho_\mathrm{DM} } d  \rho   \,, 
 \end{equation}    
 in agreement with the distribution used in Ref.~\cite{Knirck2018}.

As an analytically treatable case, we assume that the data stream was acquired on a uniform time grid. Considering the oscillating nature of the signal~(\ref{Eq:Signal}), we will work in frequency space, applying a discrete Fourier transform (DFT) to the data. 
The data set is assumed to include $N$ measurements, taken over the total observation time $T$.  To make the derivation as transparent as possible, we assume that the intrinsic instrument noise is white with variance $\sigma^{2}$.
The assumption of white noise, however, is hardly necessary, and the derivation remains valid for the more general case of colored noise. The generalization can be accomplished by replacing $N\sigma^{2}$ with the noise power spectral density $\tilde{S}(f_{p})$, where $f_p$ is the DFT frequency.   We use DFT notation and definitions of Ref.~\cite{Derevianko2018} (see appendix of that paper for a review of  Bayesian statistics in frequency space). 

Suppose we would like to establish constraints on the coupling strength $\gamma$ at one of the DFT frequencies $f_{p} = f_{\phi}$ with $\td_{p}$ being the DFT component of the original data.
A commonly used test statistic is the excess-power statistic~\cite{Groth1975} and we will use its Fourier amplitude analog, $A_p=\vert\td_{p}\rvert/\sqrt{N\sigma^2}$, see Sec.~\ref{Sec:likelihood} for a more detailed discussion of the likelihood. Note that $\theta$ becomes important when sampling time $T$ is short when compared to $1/f_c$, which we treat in Sec.~\ref{Sec:bruteforce}.
The likelihood at $f_{p}$ (we now drop the unnecessary $p$ subscript) is then
\begin{equation}
\mathcal{L}\left(A \big| A_s \right)  = 2A e^{ -
	A^2-A_s^2} 
I_{0} \left( 2AA_s\right)  \, ,
\,\label{Eq:likelihoodamplitude}%
\end{equation}
for a given excess-signal amplitude $A_s=\lvert\tilde{s}_{p}\rvert/\sqrt{N\sigma^2}$ with $\tilde{s}_p=N \gamma \xi\Phi_{0} e^{i\theta}/2$ being the DFT component of the signal~(\ref{Eq:Signal}) at $f_p=f_\phi$ and $I_0$ being the modified Bessel function of the first kind. This likelihood is general and in agreement with those used in other broadband searches~\cite{Derevianko2018,Dailey2020}. In a more general case this likelihood entails a product over the relevant frequency space bins, and over multiple channels of a network~\cite{Derevianko2018,Dailey2020,Jaynes.Spectrum.1987}.

The likelihood is the starting point of most, if not all, inference frameworks and here we will demonstrate several approaches using Eq.~\eqref{Eq:likelihoodamplitude}. We approach the problem from a Bayesian perspective and derive the posterior distribution for $\gamma$ in the stochastic and deterministic cases using both a uniform and Jeffreys' prior in Sec.~\ref{Sec:Bayesapproach}. We discuss the subtle difficulties in the use of the uniform prior for this particular problem and the benefits of utilizing objective priors. In Sec.~\ref{Sec:frequentistapproach} we demonstrate a marginal likelihood approach to compare the two cases, where the results are confirmed and extended to the pseudoscalar case via a bottom up (brute-force) Monte-Carlo (MC) simulation in Sec.~\ref{Sec:bruteforce}. 

As discussed in Refs.~\cite{Cousins1995,Heinrich2008,Lyons2008,Narsky2000,Read2000a}, the exclusion limit can depend on the chosen analysis method. We discuss several approaches in the subsequent sections. Crucially in our case, Lindley's paradox (disagreement between Bayesian and frequentist limits), is resolved with the use of the Berger-Bernardo reference prior. It is important to emphasize that, regardless of the chosen analysis, accounting for the stochastic behavior generically relaxes constraints on couplings of VULFs to standard model particles and fields.

\subsection{Bayesian approach}\label{Sec:Bayesapproach}

Bayesian inference of some parameter of interest, $y$,  requires prior information, $p(y)$, to derive the respective posterior distribution, $p(y\big\vert D)$, given the data $D$.  If there is a nuisance parameter $z$, marginalization of the joint posterior is performed to remove it using
\begin{equation}
p(y\big\vert D)=\int_{z}p(y,z\big\vert D)dz.\label{Eq:marginalization}
\end{equation}
Given the likelihood $\mathcal{L}(D\big\vert y,z)$, application of Bayes theorem yields the joint posterior

\begin{equation}
p(y,z\big\vert D)=\dfrac{p(y,z)\mathcal{L}(D\big\vert y,z)}{p(D)}=\dfrac{p(y,z)\mathcal{L}(D\big\vert y,z)}{\int_{z}\int_y p(y,z)\mathcal{L}(D\big\vert y,z)dzdy},
\label{Eq:Bayestheorem}
\end{equation}
where the denominator is a normalization constant. With the joint prior $p(y,z)=p(y)p(z\big\vert y)$ and Eq.~\eqref{Eq:marginalization}, it is equivalent to applying Bayes' theorem directly to the marginal likelihood
\begin{equation}
	p(y\big\vert D)\propto p(y)\mathcal{L}_m(D\big\vert y),
\end{equation}
with the marginal likelihood defined as 
\begin{equation}
	\mathcal{L}_m(D\big\vert y)=\int_{z}\mathcal{L}(D,z\big\vert y)dz=\int_{z}\mathcal{L}(D\big\vert y,z)p(z\big\vert y)dz.
	\label{Eq:likelihoodmarginaldef}
\end{equation}

For our particular problem, the parameter of interest is the coupling strength $\gamma$, the nuisance parameter is $\Phi_0$, and the observed data are $A$. We will compare the resulting limits on $\gamma$ using the two distributions for $\Phi_0$ shown in Eq.~\eqref{Eq:AmplDistros}. The shape of the resulting posterior distribution $p(\gamma\big\vert A)$ has immediate implications for placing constraints, also in the event of a DM detection, as the stochastic nature of the VULF affects uncertainty in the value of $\gamma$. Here, we assume that the dark matter signal is well below the detection threshold, an assumption consistent with current experimental results. Then the constraint at the 95\% credible interval (the Bayesian analog to the confidence interval), $\gamma_{95\%}$,
can be determined from
\begin{equation}
\int_{0}^{\gamma_{95\%}}  p(\gamma\big\vert A )  d\gamma = 0.95 \, .
\label{Eq:95CLdef}
\end{equation}
The correction factor obtained will be a ratio of the constraints determined from the deterministic and stochastic $\Phi_0$ distributions,  $\gamma_{95\%}^\mathrm{stoch}/\gamma_{95\%}^\mathrm{det}$. For the remainder of this section we derive constraints in terms of the excess signal amplitude $A_s$ where the results can be converted into those of $\gamma$. We note that working in terms of Fourier amplitude implies our constraint is actually on $\big\vert\gamma\big\vert$, if we worked in the complex Fourier domain the coupling could be negative and a factor of two would be introduced to Eq.~\eqref{Eq:95CLdef}.

Applying Bayes theorem, Eq.~\eqref{Eq:Bayestheorem}, to our likelihood, Eq.~\eqref{Eq:likelihoodamplitude} the posterior distribution for the excess signal amplitude, $A_s$ is
\begin{equation}
	p(A_s\big\vert A)\propto p(A_s)2A e^{ -
		A^2-A_s^2} 
	I_{0} \left( 2AA_s\right),
\end{equation}
where the exact shape of the posterior, and resulting limits, depend strongly on the choice of prior $p(A_s)$. The requirement of this prior is the main departure from frequentist or purely likelihood based approaches. While a uniform prior can seem a natural and intuitive choice for an unknown model parameter (all values are equally likely), it can lead to a number of issues that can be avoided with the use of objective priors~\cite{Gelman2017f,Berger1999a}. Objective Bayesian statistics derive prior distributions using formal rules instead of basing them off of degree of belief. Importantly, these objective priors can ensure invariance of the inference results under a change of variables (e.g. between power and amplitude), see Ref.~\cite{Kass1996} for a review. We demonstrate this lack of invariance and a resulting improper posterior when using a uniform prior after a change of variables to power in Sec.~\ref{Sec:uniformprior}.

\subsubsection{Berger-Bernardo method}\label{Sec:berger}

In this section we demonstrate the Berger-Bernardo reference prior for the case of a separation of the parameters as discussed above, with a parameter of interest and nuisance parameter $y$ and $z$ respectively~\cite{Kass1996,Bernardo1979}. First, we find the marginal likelihood as defined in Eq.~\eqref{Eq:likelihoodmarginaldef} for our particular case using the stochastic $p(\Phi_0\big\vert\Phi_{\rm{DM}})$
\begin{equation}
	\mathcal{L}_m\left(A\big\vert \Asdm\right)=\int_{\Phi_0}\mathcal{L}\left(A\big\vert  \Asdm,\Phi_0\right)p(\Phi_0\big\vert\Phi_{{\rm{DM}}})d\Phi_0=2 A e^{-A^2/(1+{\Asdm}^2)}\big/(1+{\Asdm}^2),
	\label{Eq:likelihoodmarginal}
\end{equation}
with $\Asdm=A_s\big\vert_{\Phi_0=\Phi_{\rm{DM}}}=\sqrt{N}\big\vert\gamma\xi\big\vert \Phi_{\textrm{DM}}/(2\sigma)$, the excess signal power evaluated at $\Phi_0=\Phi_{\rm{DM}}$. Note that $\mathcal{L}\left(A\big\vert\Asdm,\Phi_0\right)$ is Eq.~\eqref{Eq:likelihoodamplitude} with $A_s\rightarrow \Asdm\Phi_0/\Phi_{{\rm{DM}}}$. 

This marginal likelihood is then used to calculate the Berger-Bernardo reference prior, which happens to be equivalent (now that there are no more nuisance parameters~\cite{Bernardo1979}) to Jeffreys' prior
\begin{equation}
	p(\Asdm)\propto \sqrt{I(\Asdm)},\label{Eq:jeffreyprior}
\end{equation}
where $I(\Asdm)$ is the Fisher information,
\begin{equation}
	I(\Asdm)=-\int \mathcal{L}_m(A\big\vert \Asdm)\dfrac{\partial^2}{\partial\Asdm^2}\log \mathcal{L}_m(A\big\vert \Asdm)dA,
\end{equation}
giving the prior (please note that this is effectively a prior on $\gamma$ given the definition of $\Asdm$)
\begin{equation}
p(\Asdm)\propto \Asdm/(1+\Asdm^2).
\end{equation}
	
We can now use this prior to solve for our posterior distribution
\begin{equation}
	p_{\rm{stoch}}(\Asdm\big\vert A)=p(\Asdm)\mathcal{L}_m(A\big\vert \Asdm)/p(A)=\dfrac{2 A^2 \Asdm e^{-A^2/\left(1 + \Asdm^2\right)}}{\left(
	1 + \Asdm^2\right)^2\left(1 - e^{-A^2}\right)}.
	\label{Eq:posteriorstoch}
\end{equation}
The corresponding 95\% credible interval, Eq.~\eqref{Eq:95CLdef}, is found assuming $A$ is at the experimental detection threshold, see Sec.~\ref{Sec:frequentistapproach}, $A^{\textrm{dt}}=\sqrt{\ln{(1/\alpha)}}$ where $\alpha=0.05$ is the type-I error
\begin{equation}
	\gamma_{95\%}^\mathrm{stoch}=15.6 \dfrac{\sigma}{\sqrt{N} \xi \Phi_{{\rm{DM}}}}.
	\label{Eq:bayesStochlimit}
\end{equation}

For the deterministic case we use the other distribution of Eq.~\eqref{Eq:AmplDistros} which is equivalent to setting $A_s\rightarrow\Asdm$ in the original likelihood, Eq.~\eqref{Eq:likelihoodamplitude}. Then the deterministic posterior distribution is found using the same Berger-Bernardo reference prior (as we note below, this is technically incorrect but has little effect on the results)
\begin{equation}
	p_{\rm{det}}(\Asdm\big\vert A)\propto p(\Asdm)\mathcal{L}(A\big\vert \Asdm)\propto\dfrac{\Asdm e^{-\Asdm^2}I_0(2 A \Asdm)}{1+\Asdm^2}.\label{Eq:posteriordet}
\end{equation}
Please see Fig.~2 in the main text for a comparison of Eq.~\eqref{Eq:posteriorstoch} and Eq.~\eqref{Eq:posteriordet}.

Numerical integration under the same assumptions as before, $A=A^{\textrm{dt}}$, yields a constraint at the 95\% credible interval of
\begin{equation}
	\gamma_{95\%}^\mathrm{det}=5.28 \dfrac{\sigma}{\sqrt{N} \xi \Phi_{{\rm{DM}}}}.
	\label{Eq:bayesDetlimit}
\end{equation}

Comparing the deterministic result with Eq.~(\ref{Eq:bayesStochlimit}) derived in the stochastic treatment, we find
\begin{equation}
\gamma_{95\%}^\mathrm{stoch}  \approx 3.0  \, \gamma_{95\%}^\mathrm{det} \,. 
\label{Eq:Bayes95CLStocVsDet}
\end{equation}
We note that this factor is sensitive to the choice of $A$ and would be reduced if $A=\langle A\rangle$ rather than $A^{\textrm{dt}}$ was chosen.

Rigorously, one should calculate the Berger-Bernardo reference prior for the deterministic likelihood (which would be Jeffreys' prior, Eq.~\eqref{Eq:jeffreyprior}, using Eq.~\eqref{Eq:likelihoodamplitude} with $A_s\rightarrow\Asdm$). However, this prior does not have an analytic form and the deterministic constraint is relatively insensitive to the choice of prior. Using the aforementioned prior, the stochastic Berger-Bernardo prior, or a uniform prior all yield constraints that agree within 6.3\% of Eq.~\eqref{Eq:bayesDetlimit}. For completeness, we note that using a uniform prior, numerical integration can be avoided resulting in the deterministic posterior
\begin{equation}
p_{\rm{det}}(\Asdm\big\vert A)= \mathcal{L}(A\big\vert \Asdm)/p(A)=\dfrac{2 e^{-(A^2/2+\Asdm^2)} I_0(2 A \Asdm)}{\sqrt{\pi}I_0(A^2/2)}.
\end{equation}
 This shows that the choice between the three priors has minimal effect on the constraint when using the deterministic model for field amplitude. However, as will be clear in the next section, this insensitivity to the chosen prior is not true for the stochastic case.

\subsubsection{Stochastic case for uniform and objective priors}\label{Sec:uniformprior}

While the deterministic case is insensitive to the choice of prior, we will show that this is not true for the stochastic case. Repeating the calculation done in Eq.~\eqref{Eq:posteriorstoch} to calculate the stochastic posterior but with a uniform prior,

\begin{equation}
p_{\rm{stoch,uniform}}(\Asdm\big\vert A)\propto\mathcal{L}_m(A\big\vert \Asdm)
\label{Eq:posteriorstochuniform},
\end{equation}
yields a constraint with numerical integration of
\begin{equation}
\gamma_{95\%}^\mathrm{stoch,uniform}=63.2 \dfrac{\sigma}{\sqrt{N} \xi \Phi_{{\rm{DM}}}},
\end{equation}
and a correction factor of $\sim13$ which is considerably different from the result using the Berger-Bernardo method, Eq.~\eqref{Eq:Bayes95CLStocVsDet}.

A serious problem with the uniform prior is revealed by analyzing the effect of a change of variables on the resulting constraint. In order for the results to be meaningful and consistent, the constraint should be independent of the choice to work in amplitude or power (as is the case in the frequentist approach discussed in the next section). In particular, using the uniform prior after a change of variables to excess power, $P=\lvert\tilde{d}_{p}\rvert^2/(N\sigma^2)$,
\begin{equation}
p_{\rm{stoch,uniform}}(P_s^{\textrm{DM}}\big\vert P)=\dfrac{\mathcal{L}_m(P\big\vert P_s^{\textrm{DM}})}{\int\mathcal{L}_m(P\big\vert P_s^{\textrm{DM}})dP_s^{\textrm{DM}}}
\label{Eq:posteriorpowerstochuniform},\quad \textrm{with}\quad \mathcal{L}_m(P\big\vert P_s^{\textrm{DM}})=e^{-P/(1+P_s^{\textrm{DM}})}/(1+P_s^{\textrm{DM}})
\end{equation}
reveals a denominator that does not converge, yielding an improper posterior distribution. See Eq.~\eqref{Eq:likelihoodpower} in Sec.~\ref{Sec:likelihood} for the likelihood in terms of excess power $P$. Use of an objective prior addresses exactly this issue. Using the Berger-Bernardo method shown above, Eq.~\eqref{Eq:jeffreyprior}, in the context of the excess-power test statistic gives the prior
\begin{equation}
	p(P_s^{\textrm{DM}})=1/(1+P_s^{\textrm{DM}}),
\end{equation}
where we again note that this is effectively a prior on $\gamma$ with $P_s^{\textrm{DM}}=P_s\big\vert_{\Phi_0=\Phi_{\rm{DM}}}=N\gamma^2\xi^2 \Phi_{\textrm{DM}}^2/(4\sigma^2)$.

Using this prior yields the constraint
\begin{equation}
\gamma_{95\%}^\mathrm{stoch}= 15.6 \dfrac{\sigma}{\sqrt{N} \xi \Phi_{{\rm{DM}}}},
\end{equation}
agreeing within a fraction of a percent with the result obtained for the analysis using amplitude. For this reason, we find that the effect of the stochastic nature of the VULF on experimental constraints is most reasonably estimated in the Bayesian framework by using the objective prior. Furthermore, as we show in the next section, the effect of stochastic fluctuations of the VULF on constraints analyzed using a frequentist approach yields results similar to that obtained in the Bayesian framework employing the objective prior in Sec.~\ref{Sec:berger}.

\subsection{Frequentist approach using marginal likelihood}\label{Sec:frequentistapproach}
\begin{figure}[h!]
	\includegraphics[scale=1]{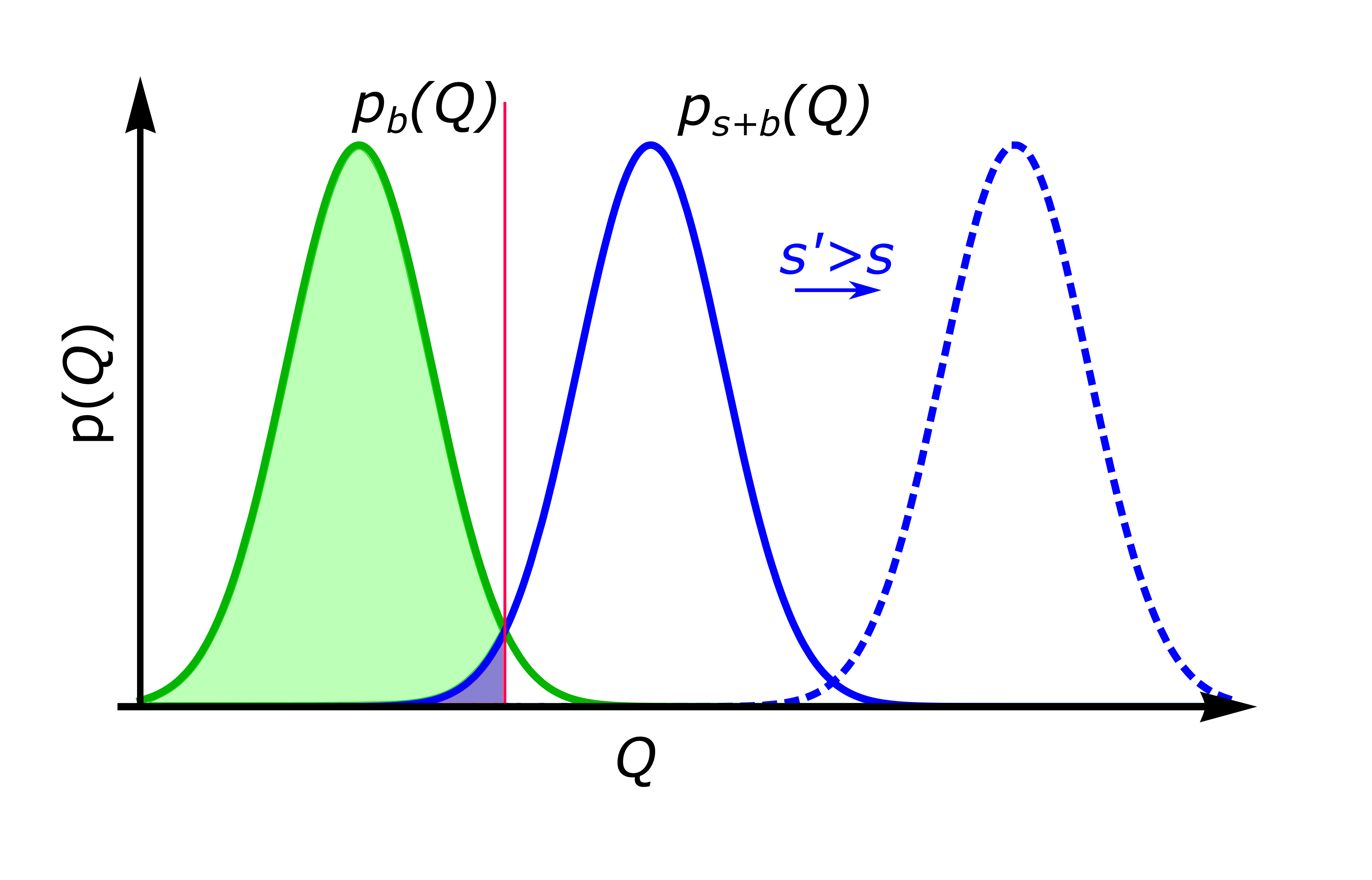}
	\caption{Illustrative figure of a hypothesis test for background vs. signal plus background. The figure represents PDFs of the test statistic $Q$ for various values of the signal parameter. The green curve represents the background only hypothesis and the blue curves signal values $s>0$. The Type-I error, $\alpha$, and detection threshold, red line, are set by the chosen confidence level, $CL=1-\alpha$, which is represented by the green area of the background only hypothesis $p_b(Q)$. The blue area represents the Type-II error, $1-\beta$, and meets the condition $1-\beta=\alpha$ satisfied at the exclusion threshold, which is $s$ for this figure. Signals stronger than the threshold $s'>s$ represent models within the exclusion region, dashed-blue lines.}
	\label{Fig:freqexclusion}
\end{figure}

A frequentist constraint can be obtained via hypothesis testing and using the Type-I error (or false alarm rate), $\alpha$, and Type-II error (or false detection rate), $1-\beta$ (see Ref.~\cite{kendall1973advanced} or more recent editions). Type-I error is the rejection rate of the null hypothesis when it is true while the Type-II error is the acceptance (or non-rejection) of the null hypothesis when an alternative (signal) hypothesis is true. We illustrate a critical (exclusion) region and threshold that matches these two errors in Fig.~\ref{Fig:freqexclusion}.

Our background only hypothesis is the likelihood, Eq.~\eqref{Eq:likelihoodamplitude}, with $A_s=0$ giving a detection threshold at a confidence level $CL=1-\alpha$
\begin{equation}
	1-\alpha=\int_{0}^{A^{\textrm{dt}}}\mathcal{L}\left(A \big| A_s=0 \right) dA =\int_{0}^{A^{\textrm{dt}}} 2A e^{ -A^2}dA \quad\implies\quad A^{\textrm{dt}}=\sqrt{\ln{(1/\alpha)}}.
\end{equation}

The goal is to ensure that false exclusion of the true signal value, $1-\beta$, does not occur more often than the Type-I error, $1-\beta\leq\alpha$, for any signal strength above a certain threshold. The threshold signal strength, $A_s^{\rm{th}}$, at $1-\beta=\alpha$ determines the constraint at a confidence $CL$ (or $\lesssim\alpha$ for discontinuous distributions). So, for the signal plus background case
\begin{equation}
	1-\beta=\int_{0}^{A^{\textrm{dt}}}\mathcal{L}\left(A \big| A_s=A_s^{\rm{th}} \right) dA,
\end{equation}
which yields a limit at the 95\% CL ($\alpha=0.05$) in terms of $\gamma$ of
\begin{equation}
	\gamma_{95\%}^{\rm{det}}=5.6\dfrac{\sigma}{\sqrt{N} \xi \Phi_{{\rm{DM}}}}.
\end{equation}
This limit agrees with the Bayesian approach within 5\% using the Berger-Bernardo prior and within $1\%$ using the uniform prior.

To treat the stochastic case we repeat the above calculation using the marginal likelihood, Eq.~\eqref{Eq:likelihoodmarginal}. There is some debate behind treating the likelihood as a PDF in this fashion~\cite{Berger1999a}, but it is occasionally done in frequentist approaches~\cite{Tanabashi2018} and produces reasonable results in our case.

Repeating the previous derivation replacing $\mathcal{L}$ with $\mathcal{L}_m$ yields the constraint
\begin{equation}
	\gamma_{95\%}^{\rm{stoch}}=15.1\dfrac{\sigma}{\sqrt{N} \xi \Phi_{{\rm{DM}}}}.
\end{equation}
This result is within $3\%$ of the Bayesian approach with the Berger-Bernado prior, but is $\sim4$ times smaller than that using a uniform prior. This constraint agrees with a brute-force MC, Sec.~\ref{Sec:bruteforce}, including reasonable fits to the distributions used in this section.

\subsection{Additional stochastic parameters for the pseudoscalar coupling and brute-force MC}\label{Sec:bruteforce}
The gradient coupling of a bosonic field introduces an additional factor to the signal~\eqref{Eq:Signal} proportional to $\vec{v}\cdot\vec{e}$ where $\vec{v}$ is the DM velocity in the lab frame and $\vec{e}$ is the sensitive axis of the experiment, see discussion in Refs.~\cite{wu2019search,garcon2019constraints}. The velocity, $\vec{v}$, can be written in terms of separate contributions with the leading term being the Sun's velocity toward Cygnus, additional modulation due to Earth's rotation (including orbit), and the random contribution of the field's velocity. We assume full modulation from the Earth's rotation, i.e the detector rotates parallel and perpendicular to the Cygnus component daily for a detector with sensitive axis $\hat{e}=\hat{z}$. These approximations give the expression $\vec{v}\cdot\vec{e}\approx v\cos(2\pi f_Et)$ where $f_E=1/\text{day}$. We use a Gaussian distribution for the field velocity with mean value and spread of $10^{-3}c$, where we assume the Sun's speed is the same. This allows us to write a new definition of the signal as
\begin{equation}
s(t)=\gamma\xi\Phi_0 m_\phi \cos(2\pi f_\phi t+\theta)\vec{v}\cdot\vec{e}\approx\gamma\xi v\Phi_0m_\phi \cos(2\pi f_\phi t+\theta)\cos(2\pi f_E t).
\label{Eq:SignalWind}
\end{equation}
This expression naturally defines three regimes: writing down the coherent oscillation assumes we are in the middle regime of $T\ll\tau_c$ and the third regime occurs when the daily modulation dominates for $1/f_E<T\ll1/f_\phi$. We use this model to calculate the correction in the $T<1/f_\phi$ regime for the Cosmic Axion Spin Precession Experiment with a Zero to Ultra-low Field Comagnetometer, CASPEr-ZULF-Comagnetometer, data presented in Ref.~\cite{wu2019search}, and to confirm the results in the previous sections.

In short, we compare MC excess signal power thresholds under the two signal assumptions (stochastic vs. deterministic parameters) while scanning the signal frequency in the regime $T<1/f_\phi$. To do so we generate 10$^6$ MC data for each frequency $f_\phi$, choosing $v$, $\theta$, and $\Phi_0$ from their respective distributions, sampling for a total time $T$ of 30 days at a rate above a Nyquist frequency set by $f_E$. The threshold coupling, $\gamma_{95\%}$, guarantees that 95\% of the excess amplitudes at that particular frequency lie above the background only detection threshold. This ensures the condition that $\alpha=1-\beta$ as discussed in Sec.~\ref{Sec:frequentistapproach}. These thresholds are compared to those found using the deterministic data, generated by mirroring the assumptions made in the original paper~\cite{wu2019search}, where the ratio between the deterministic and stochastic thresholds provides the correction factor to be applied to the original exclusion plot. We note that this MC approach is similar to the CL$_s$ method~\cite{Read2000a}. The results of this simulation yields the roll-off seen for the CASPEr-ZULF-Comagnetometer data in Fig.~\ref{Fig:axionexclusion} for masses below $\sim10^{-21}$ eV.

\begin{figure}[h!]
\begin{subfigure}{.49\linewidth}
	\centering
	\includegraphics[width=\linewidth]{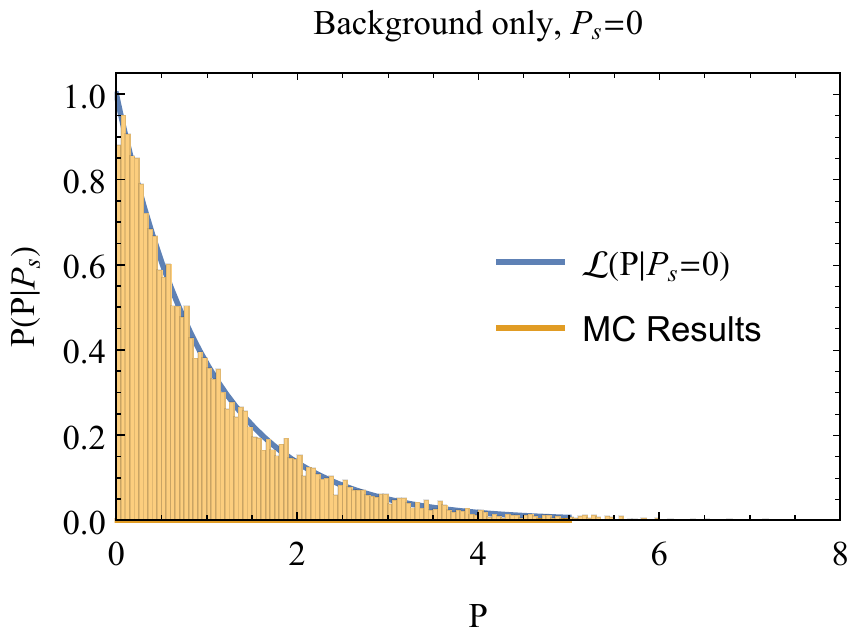}
	\caption{}
	\label{fig:sfig1a}
\end{subfigure}
\begin{subfigure}{.49\linewidth}
	\centering
	\includegraphics[width=\linewidth]{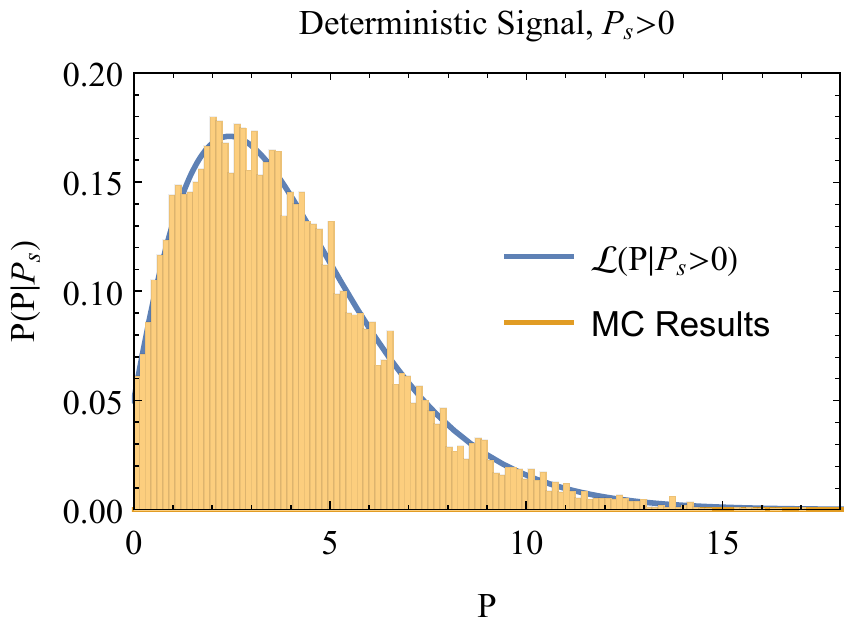}
	\caption{}
	\label{fig:sfig1b}
\end{subfigure}
\begin{subfigure}{.49\linewidth}
	\centering
	\includegraphics[width=\linewidth]{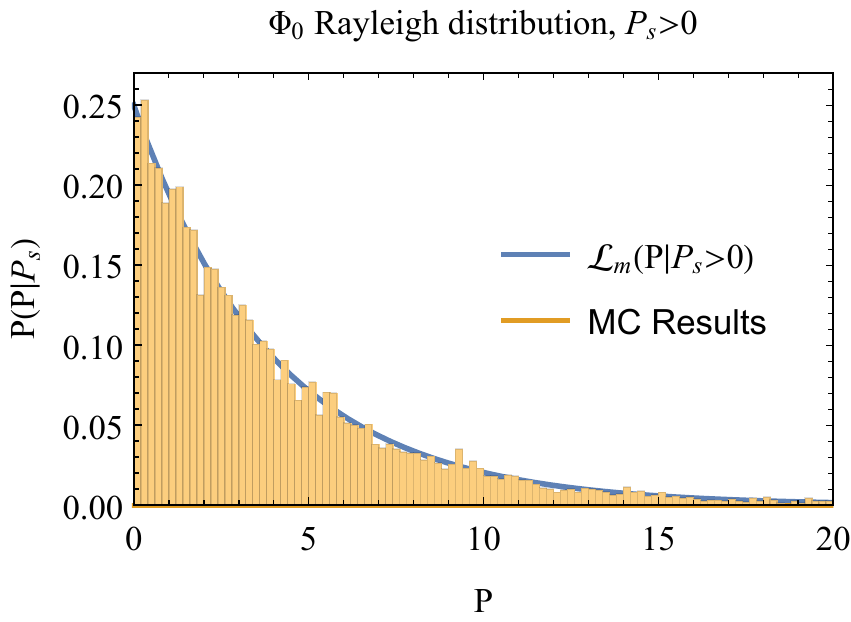}
	\caption{}
	\label{fig:sfig1c}
\end{subfigure}
\begin{subfigure}{.49\linewidth}
	\centering
	\includegraphics[width=\linewidth]{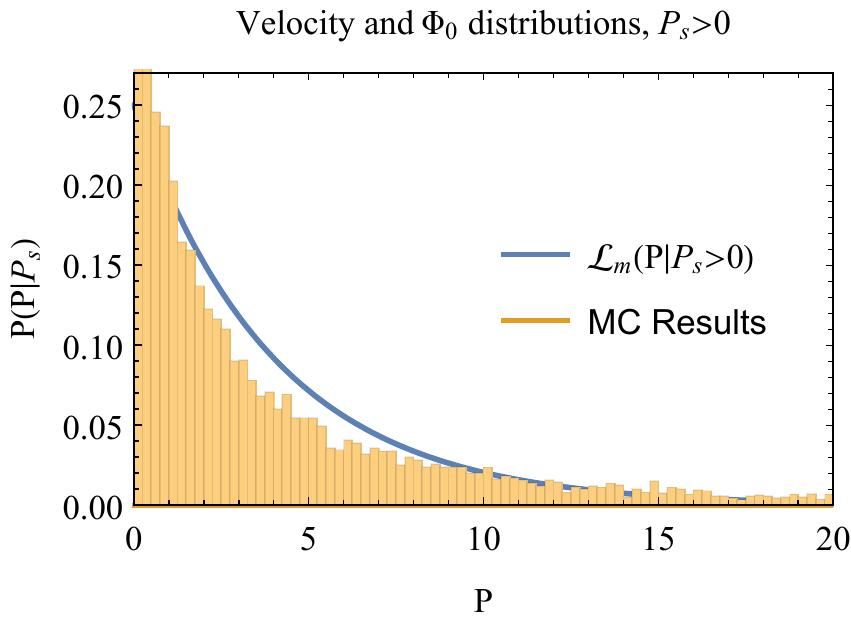}
	\caption{}
	\label{fig:sfigd}
\end{subfigure}
\caption{Results of the excess-power statistic, $P$, at the signal frequency, $f_\phi$, for 10,000 MC signals under various options of the signal parameters. (a) Background only model with $P_s=0$ showing expected exponential. (b) Deterministic signal model showing non-central chi-square distribution as shown by Eq.~\eqref{Eq:likelihoodpower}. (c) With a Rayleigh distributed $\Phi_0$ the MC results are fit by the marginal likelihood, Eq.~\eqref{Eq:posteriorpowerstochuniform}. (d) The change induced from the marginal likelihood seen in (c) after turning on the velocity distribution, revealing that lower powers are now more likely at the same coupling strength.}
\label{Fig:MCdistributions}
\end{figure}

\begin{figure}[h!]
	\includegraphics[width=.49\linewidth]{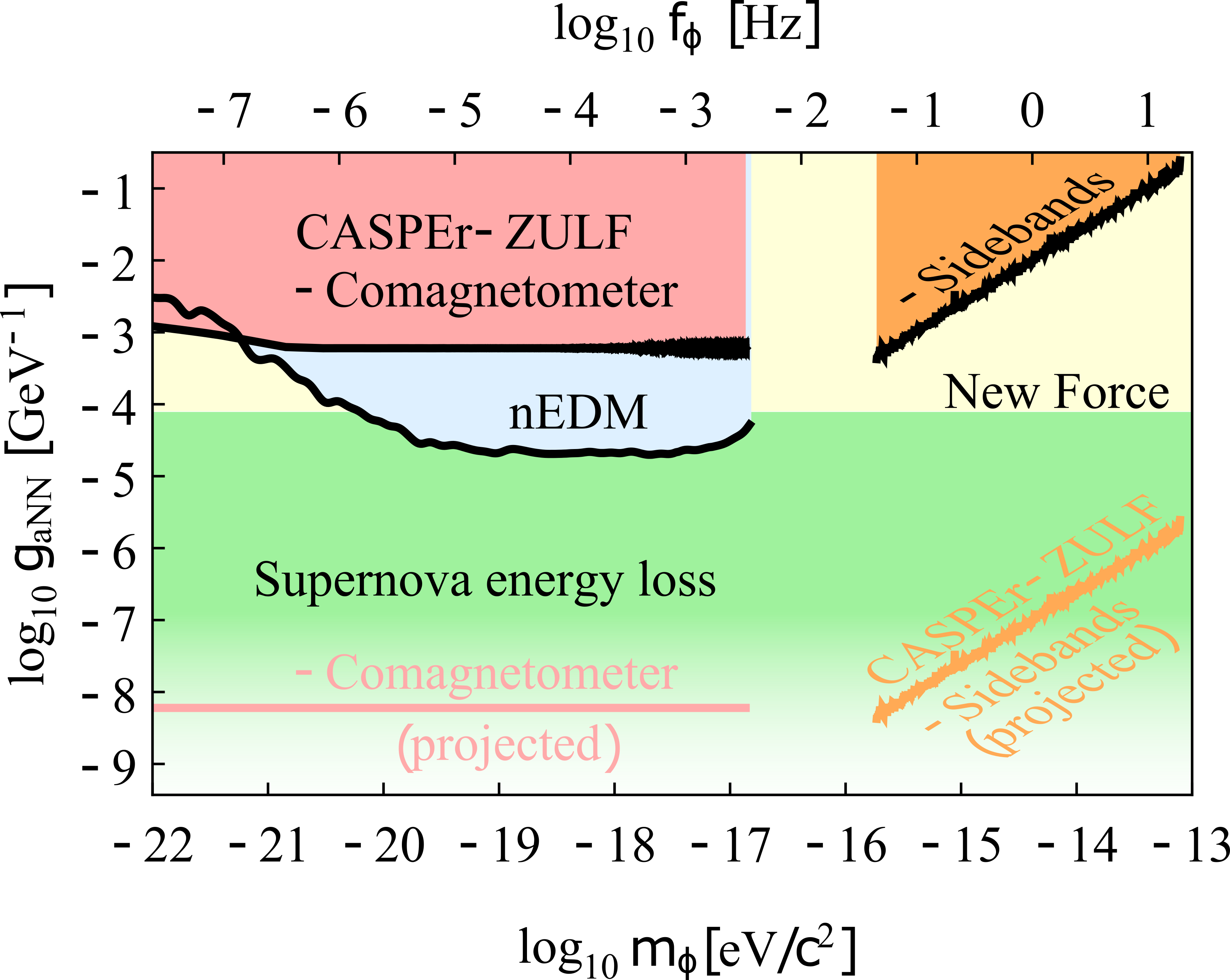}
	\caption{An updated VULF exclusion plot of laboratory constraints for the axion-nucleon coupling, $g_{aNN}$, from the following experiments: CASPEr-ZULF~\cite{wu2019search,garcon2019constraints}, nEDM~\cite{Abel2017}, and a new-force search using a K-$^3$He comagnetometer~\cite{Vasilakis2009}.}
	\label{Fig:axionexclusion}
\end{figure}

We also used this MC to confirm the results in the previous section and found that the distributions for the deterministic case, Eq.~\eqref{Eq:likelihoodpower}, and with a Rayleigh distributed field amplitude, Eq.~\eqref{Eq:posteriorpowerstochuniform}, were reproduced exactly, see (a)-(c) of Fig.~\ref{Fig:MCdistributions}. 

We used this MC to determine the correction factor in the middle regime, where the $\Phi_0$ and $v$ distributions are relevant, see (d) in Fig.~\ref{Fig:MCdistributions}, yielding
\begin{equation}
\gamma_{95\%}^\mathrm{stoch}  = 8.4  \, \gamma_{95\%}^\mathrm{det} \,,
\label{Eq:FreqPseudoscalarfactor}
\end{equation}
which was used to adjust the limits presented in Fig.~\ref{Fig:axionexclusion}. This result is similar to that obtained in Ref.~\cite{garcon2019constraints}, where the actual experimental projection $\vec{v}\cdot\vec{e}$ was taken into account. Since Eq.~\eqref{Eq:SignalWind} changes depending on said projection, $\vec{v}\cdot\vec{e}$, for a particular experiment, the factor may no longer be $\sim\!8.4$ and should be calculated case by case. We acknowledge the fact that the variation of the three parameters $\Phi_0$, $v$, and $\theta$ discussed here may be correlated, and that a more thorough theoretical treatment may provide some corrections. Nonetheless, the results presented here support the argument that stochastic variation is essential to consider when constructing exclusion plots in the appropriate regimes, or in the event of a detection.

\subsubsection{Constraints below frequency resolution ($1/T$)}
In this section we discuss marginalization over the unknown phase $\theta$ of the VULF and, in particular, the effect of this marginalization on searches for VULFs with frequencies below $1/T$. This problem was the subject of a recent debate in the literature related to the CASPEr-ZULF-Comagnetometer limits originally presented in Ref.~\cite{wu2019search}. The comment~\cite{Adelberger2019Comment} on the published article~\cite{wu2019search} correctly pointed out that the limits for frequencies below the $1/T$ frequency resolution should not improve compared to limits for frequencies above the $1/T$ frequency resolution. This is true and the mistake in Ref.~\cite{wu2019search} came from not marginalizing over the uniform distribution of phase, $\theta$ in Eq.~\eqref{Eq:SignalWind} above, and erroneously assuming the average signal amplitude in this limit $\propto\frac{1}{2\pi}\int\lvert\sin(\theta)\rvert d\theta=2/\pi$, a deterministic approach to $\theta$ similar in principle to the erroneous deterministic approaches to VULF amplitudes and velocities~\cite{Wu2019Reply}. However, after accounting for this unknown phase as discussed both in the present work and in the reply~\cite{Wu2019Reply}, the authors of the comment~\cite{Adelberger2019Comment} still disagreed with the corrected limit published here. 

The resolution of this debate is rooted in how constraints on couplings of VULF dark matter to standard model particles and fields should be interpreted. Even when assuming that VULF dark matter is well described by the SHM, at a particular time and place, the experimentally observable signal is affected by a number of unknown VULF parameters beyond the coupling strength such as the phase, velocity, and amplitude. If the duration $T$ of an experiment is too short to broadly sample the distribution of the unknown VULF parameters, then in actuality the experiment excludes the existence of VULF dark matter over some finite volume of a multi-dimensional space of the unknown parameters. To condense the results of a VULF dark matter search into a two-dimensional plot of coupling strength vs. frequency, it is necessary to marginalize over the other unknown VULF parameters. While the authors of the comment~\cite{Adelberger2019Comment} object to this marginalization, arguing that for particular phases $\theta$ the constraints on couplings could be much weaker, we argue that it is reasonable to plot constraints on couplings ruled out over 95\% of the multi-dimensional unknown VULF parameter space. Furthermore, as we note here, the authors of the comment~\cite{Adelberger2019Comment} have treated the amplitude and velocity of the VULF dark matter as deterministic in their own experiment~\cite{terrano2019constraints}. For particular (but unlikely) VULF velocities and amplitudes, the experiment~\cite{terrano2019constraints} is insensitive to VULF couplings. In order to constrain VULF dark matter couplings in the regime where $T$ is shorter than either the coherence time or period of the VULF, it is necessary to carry out some form of marginalization over the unknown VULF parameters.

The derived constraints on VULF dark matter presented in Ref.~\cite{wu2019search} take advantage of the signal model presented in Eq.~\eqref{Eq:SignalWind}. Any pseudoscalar DM signal with frequency $f_\phi$ below the $1/T$ resolution will be upconverted to the 1/day frequency ($f_E$) given by Earth's rotation. Thus, instead of trying to resolve frequencies below the $1/T$ resolution, we constrain all frequencies below that threshold by looking only at the $f_E$ bin that is within the detection bandwidth. The constraint levels off (becomes constant) for all frequencies below $\sim1/(25T)$ when $2\pi f_\phi T \approx 0$ and the signal amplitude is dominated by the $\lvert\sin\theta\rvert$ distribution. The height of this constant level is determined by the chosen confidence level, as there are a small range of phases that give vanishing signals for arbitrarily large coupling. 

This approach can be used for all pseudoscalar (``wind"-type) experiments that can resolve the $f_E$ frequency. We additionally note that scalar searches (for which this upconversion does not occur) which are sensitive to offsets (the zero-frequency component) can also draw similar constraints below the $1/T$ frequency threshold. There is a caveat that any signal present at the $f_E$ frequency would not allow determination of the actual $f_\phi$, only that it is smaller than $1/T$. Additionally, differentiating systematic errors from potential VULF dark matter signals is difficult for upconverted frequencies; one approach would be to confirm the expected yearly modulation in the event of a detection.

\subsection{Blueice Monte Carlo}

Here we employ a method widely used in astroparticle physics~\cite{Conrad2015} to compare to the Bayesian and frequentist approaches described in the previous sections. In the case encountered here we find that Wilks theorem, the use of a chi-squared distribution for the log likelihood ratio test statistic, does not hold. The reason for this is in part due to the fact that we are not using the profile likelihood but marginalize over the amplitude fluctuation by drawing from the Rayleigh distribution in each MC realization. In a situation like this, one has to resort to doing a so-called Neyman construction, see e.g. discussion in Ref.~\cite{Tanabashi2018}.

The decisive feature of a Neyman construction is that it provides the statistically desired properties (in this case the coverage of the confidence intervals). It uses the distribution of the test statistic as estimated by many MC simulations instead of the asymptotic distribution used in Wilks theorem. For a given signal strength $\gamma$, the test-statistic distribution is obtained by running many MC simulations at that signal strength. The critical value is the value of the test statistic at the $90$th (or $95$th) percentile of the distribution for a test at the $90\%$ (or $95\%$) confidence level.
This procedure is repeated for different signal strengths such that a critical value curve as a function of the signal strength is obtained.
The upper limit on the signal strength for a given measurement (or MC simulation) is then given by the value at which the likelihood crosses the critical value curve.

The correction factor is finally not determined from a single experiment but rather from an ensemble of experiments over which the so called sensitivity (median limit in case of no signal) is calculated. The correction factor is then defined analogous to the case of the Bayesian analysis as:
\begin{equation}
correction = \frac{\text{median}(\boldsymbol{u}_{stochastic})}{\text{median}(\boldsymbol{u}_{deterministic})},
\end{equation}
where $\boldsymbol{u}$ is a large number of upper limits at a certain confidence level. Initial results from the simulations show a correction factor of $4.4$ at the 95$\%$ confidence level when including only the field amplitude distribution, see Sec. 6 of Ref.~\cite{Pelssers2020} for simulation details.

\section{Modified constraint plots}
Here we present modified parameter exclusion plots for dark matter coupling constants where we apply a correction factor based on the analysis method used by the respective experiment.

\subsection{Pseudoscalar coupling}
The combined corrections for the ``wind-type" experiments CASPEr-ZULF-Comagnetometer~\cite{wu2019search}, -Sideband~\cite{garcon2019constraints}, and nEDM (neutron Electric Dipole Moment measurement)~\cite{Abel2017} account for the stochastic parameters: amplitude, relative velocity between the field and the spin~\cite{Graham2015}, and initial field phase when appropriate (for $T\ll \tau_c$ and additionally $T\ll1/f_\phi$ for phase).

We use the results of Sec.~\ref{Sec:bruteforce} that show constraints are a factor of $\sim\!8.4$ weaker than those derived assuming the deterministic parameters.
The nEDM and CASPEr-ZULF-Comagnetometer data were corrected by this factor, and we note that the published CASPEr-ZULF-Sidebands data took these parameters into account~\cite{garcon2019constraints} but made an incorrect approximation~\footnote{The scale parameter for the velocity distribution was off by a factor of $\sqrt{2}$ to correctly reproduce the Maxwell Boltzmann speed distribution. Additionally, implementing the distributions directly onto the originally derived constraint is equivalent to assuming a large signal limit and the stated confidence levels do not match the coverage criteria discussed in Sec.~\ref{Sec:frequentistapproach}.}. The published limits of CASPEr-ZULF-Sidebands used a 90\% CL, so we adjusted these data to a $95\%$ CL for consistency using the factor determined from the ratio $\gamma_{95\%}^\mathrm{det}/\gamma_{90\%}^\mathrm{det}$. The roll-off starting at $f_\phi\sim10^{-21}$ for the CASPEr-ZULF-Comagnetometer limits is due to the additional phase sensitivity, further discussed in Sec. E. The factor of $\sim\!8.4$ reduces in magnitude with decreasing projection of the detector sensitivity axis onto the lab-frame velocity.

\subsection{Scalar coupling}
The scalar coupling has only the field amplitude parameter to account for in the correction factor as treated in the main text. We use the result from the marginal likelihood of $\gamma_{95\%}^\mathrm{stoch}/\gamma_{95\%}^\mathrm{det}=2.7$ from Sec.~\ref{Sec:frequentistapproach} for the experiments using a frequentist based analysis and 3.0 from Eq.~\eqref{Eq:Bayes95CLStocVsDet} for the experiment using a Bayesian framework~\cite{Hees2016}.

\begin{figure}[h!]
	\includegraphics[width=.49\linewidth]{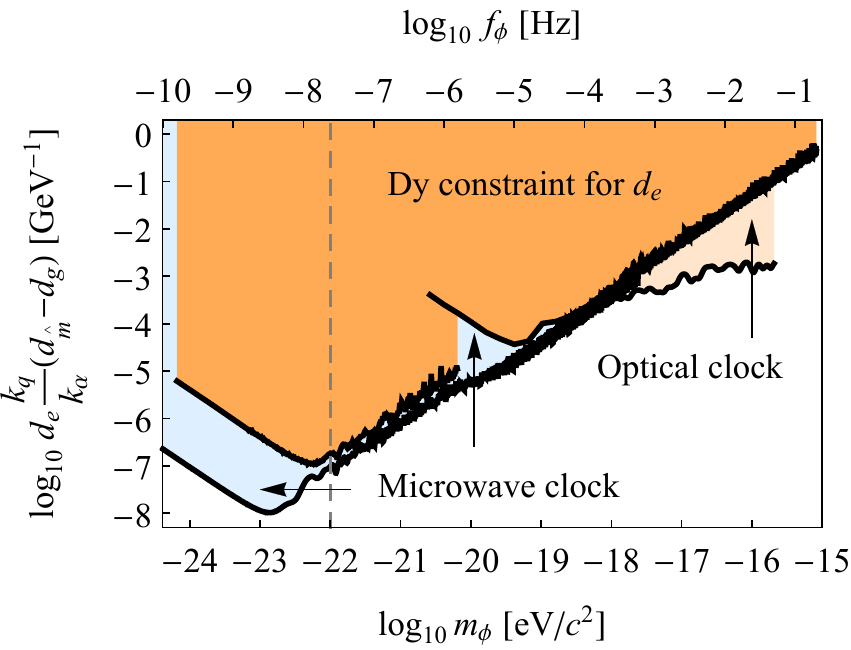}
	\caption{An updated VULF exclusion plot of laboratory constraints for dilaton coupling strength $d_e$ and linear combination (seen on y-axis~\cite{Hees2016}). Details on the data for the dual rubidium and cesium cold atom fountain FO2 at LNE-SYRTE can be found in Ref.~\cite{Hees2016}, for the dysprosium spectroscopy in Ref.~\cite{VanTilburg2015}, and from the global network of optical atomic clocks in Ref~\cite{Wciso2018}. The gray dashed line, $m_\phi\lesssim10^{-22}$ eV, indicates where the assumption that the field with mass $m_\phi$ makes up all the dark matter should be relaxed~\cite{Marsh2016}.}
	\label{Fig:dilatonexclusion}
\end{figure}
\pagebreak
\section{Likelihood}\label{Sec:likelihood}

Here we demonstrate the change of notation from Ref.~\cite{Derevianko2018} to that used in this paper, and discuss the change of parameters between complex Fourier components, amplitude, and power. Starting with the signal, Eq.~\eqref{Eq:Signal}, and with $\td_{p}$ the DFT component of the original data, the likelihood at $f_{p}$ is given by
\begin{equation}
\mathcal{L}\left( \td_{p}\big\vert \Phi_{0}, \gamma, \theta \right)  = \frac
{1}{\pi N \sigma^{2} } \exp\left\{ -
\frac{\left\vert \td_{p} - \tilde{s}_{p} \right\vert ^{2}}{N\sigma^{2}} \right\}
\,,\label{Eq:deterministic0}%
\end{equation}
where $\tilde{s}_{p} = N \gamma \xi\Phi_{0} e^{i\theta}/2$ is the DFT component of the signal~(\ref{Eq:Signal}). See Appendix of Ref.~\cite{Derevianko2018} for a review. For the phase, one can marginalize over $\theta$ as done in Ref.~\cite{Jaynes1987}, however the resulting likelihood can only be used for Bayesian approaches, see discussion in Ch.~2 and Appendix A of Ref.~\cite{Bretthorst1988}. To remain consistent with frequentist based approaches we can instead do a change of variables to excess amplitude $A=\big\vert\tilde{d}_{p}\big\vert/\sqrt{N\sigma^2}$. First we rewrite Eq.~\eqref{Eq:deterministic0} in polar coordinates using $\td_{p}=\lvert \td_{p}\rvert\exp^{-i\phi}$ and $d\td_{p}=\lvert \td_{p}\rvert d\lvert \td_{p}\rvert d\phi$ as

\begin{equation}
\mathcal{L}\left( \lvert \td_{p}\rvert,\phi\bigg\vert \Phi_{0}, \gamma, \theta \right)  = \frac
{\lvert \td_{p}\rvert}{\pi N \sigma^{2} } \exp\left\{ -
\frac{\lvert \td_{p}\rvert^2+\lvert \tilde{s}_{p} \rvert^2}{N\sigma^{2}} \right\}\exp\left\{ 
\frac{2\lvert \td_{p}\rvert\lvert \tilde{s}_{p} \rvert\cos(\phi-\theta)}{N\sigma^{2}} \right\}
\,,\label{Eq:deterministicpolar}%
\end{equation}
and then do a vector to scalar, $(\lvert \td_{p}\rvert,\phi)\rightarrow A$, change of variables
\begin{eqnarray}
	\mathcal{L}\left(A \big| A_s \right)  &=&\int d\lvert \td_{p}\rvert \int d\phi ~\delta\left(A-\dfrac{\lvert \td_{p}\rvert}{\sqrt{N\sigma^2}}\right)\mathcal{L}\left( \lvert \td_{p}\rvert,\phi\big\vert \Phi_{0}, \gamma, \theta \right)\\
	&=&\int d\lvert \td_{p}\rvert \int d\phi ~\sqrt{N\sigma^2}~\delta\left(A\sqrt{N\sigma^2}-\lvert \td_{p}\rvert\right)\mathcal{L}\left( \lvert \td_{p}\rvert,\phi\big\vert \Phi_{0}, \gamma, \theta \right) \\
	&=& 2A e^{ -
		A^2-A_s^2} 
	I_{0} \left( 2AA_s\right)  \, ,
	\,
\end{eqnarray}
with excess signal amplitude $A_s=\lvert\tilde{s}_{p}\rvert/\sqrt{N\sigma^2}$ and $I_0$ is the modified Bessel function of the first kind.

It is also common to work with the power spectral density~\cite{Groth1975} and excess power, $P=\lvert\tilde{d}_{p}\rvert^2/(N\sigma^2)$, where another change of variables yields
\begin{equation}
	\mathcal{L}\left(P \big| P_s \right)  = e^{ -(P+P_s)} 
I_{0} \left( 2\sqrt{PP_s}\right)  \, ,
\,
\label{Eq:likelihoodpower}
\end{equation}
with $P_s$ defined accordingly. See Ref.~\cite{Groth1975} for the original and alternate derivation of Eq.~\eqref{Eq:likelihoodpower}.

\section{Coherence Time}
  Another topic is the presence of various definitions of coherence time used throughout the literature. The most common definition used is $\tau_{c}\equiv\left(f_{c} v_{\mathrm{vir}}^{2}/c^{2} \right)^{-1}$ and is the one used in this paper. It is a lower bound derived by considering our motion through the spatial gradients of the field~\cite{Graham2013a,Budker2014} given by $\lambda$, the De Broglie wavelength,
  \begin{equation}
  \tau_c\sim\lambda/v_{\mathrm{vir}}=\frac{h}{m_\phi v_{\mathrm{vir}}^2}=\left(f_{c} v_{\mathrm{vir}}^{2}/c^{2} \right)^{-1}\approx10^6/f_c=1s\left(\frac{\text{MHz}}{m_\phi}\right),
  \end{equation}
  where we have used the Compton frequency $f_c=m_\phi c^2/h$ and $v_{\mathrm{vir}}\approx10^{-3}c$. However, some definitions differ by factors around $\sim2\pi$~\cite{Derevianko2018}. A factor of 2 can occur from using the kinetic energy $mv^2/2$ instead of $mv^2$ as in Ref.~\cite{Graham2013a,Budker2014} or in using 10$^6$ as the quality factor and corresponding definition of coherence time.
  
  These differences in coherence time do not have a significant effect on the results in this paper nor previous published constraints. The strongest effect would be the shift of the upper bound on $m_a$ of the CASPEr-ZULF-Sidebands constraint. Since the experiment is incoherently averaging past this point the constraint significantly weakens and was not reported. In addition, the fuzzy definition of coherence time also depends on the relative velocity, making it a distributed quantity.
  
  Regarding this upper bound, a potentially more precise $T_{\textrm{max}}$ rather than the coherence time can be derived (where the regime discussed in this paper would go from $T\ll\tau_c$ to $T\ll T_{\textrm{max}}$) considering the lineshape derived in Ref.~\cite{Derevianko2018}. Using the derived full width at half maximum $\Delta\omega\approx 2.5/\tau_c$ and setting it equal to the DFT frequency step $\delta\omega=2\pi/T$ gives $T_{\textrm{max}}\approx2\pi\tau_c/2.5$, or with the definition of coherence time used in this paper: $T_{\textrm{max}}\approx\tau_c/2.5$. Thus it is safe to assume the data from the experiments shown are all in the $T\ll T_{\textrm{max}}$ regime, regardless of the definition of coherence time.
  \section{Rayleigh Distribution}
  As shown in Eq.~\eqref{Eq:Signal}, $\Phi_0$ is the observed field amplitude, and in the case of total interrogation time being much less than the field's coherence time, $T\ll\tau_c$, the only field amplitude observed during a measurement. One can relate the local dark matter density $\rho_{\text{DM}}$ to the field's energy density $\Phi_0^2f_\phi^2/2=\rho_{\text{DM}}$ as done for example in Ref.~\cite{Graham2013a}. The problem with this is that $\Phi_0$ is actually a Rayleigh distributed value, thus the correct expression is
  \begin{equation}
  \langle\Phi_0^2\rangle \frac{c^2m_\phi^2}{2\hbar^2}=\rho_{\text{DM}}\label{eqenergydensity} \,.
  \end{equation} 
  
  We can define $\Phi_{\text{DM}}$ as the value satisfying Eq. \eqref{eqenergydensity}
  \begin{equation}
  \Phi_{\text{DM}}=\sqrt{\langle\Phi_0^2\rangle}=\frac{\sqrt{2\rho_{\text{DM}}}}{m_\phi},
  \end{equation} 
  analogous to previous assumptions of $\Phi_0=\Phi_{\text{DM}}$ in the deterministic approach.
  
  Using Eq.~\eqref{eqenergydensity} we can solve for the required $\sigma$ of the Rayleigh distribution $p(\Phi_0)$.
  
  \begin{eqnarray}
  \langle\Phi_0^2\rangle=\int\Phi_0^2 p(\Phi_0) d \Phi_{0}=&&\int_{0}^{\infty}\frac{\Phi_0^3}{\sigma^2}\exp\left(-\frac{\Phi_0^2}{2\sigma^2}\right)d \Phi_{0}=\Phi_{\text{DM}}^2\nonumber\\
  \implies &&\sigma=\Phi_{\text{DM}}/\sqrt{2}\text{ thus, }\nonumber\\
  p(\Phi_0)&&=2\frac{\Phi_0}{\Phi_{\text{DM}}^2}\exp\left(-\frac{\Phi_0^2}{\Phi_{\text{DM}}^2}\right).\label{phi0dist}
  \end{eqnarray}

\begin{figure}[!h]
	\includegraphics[scale=.6]{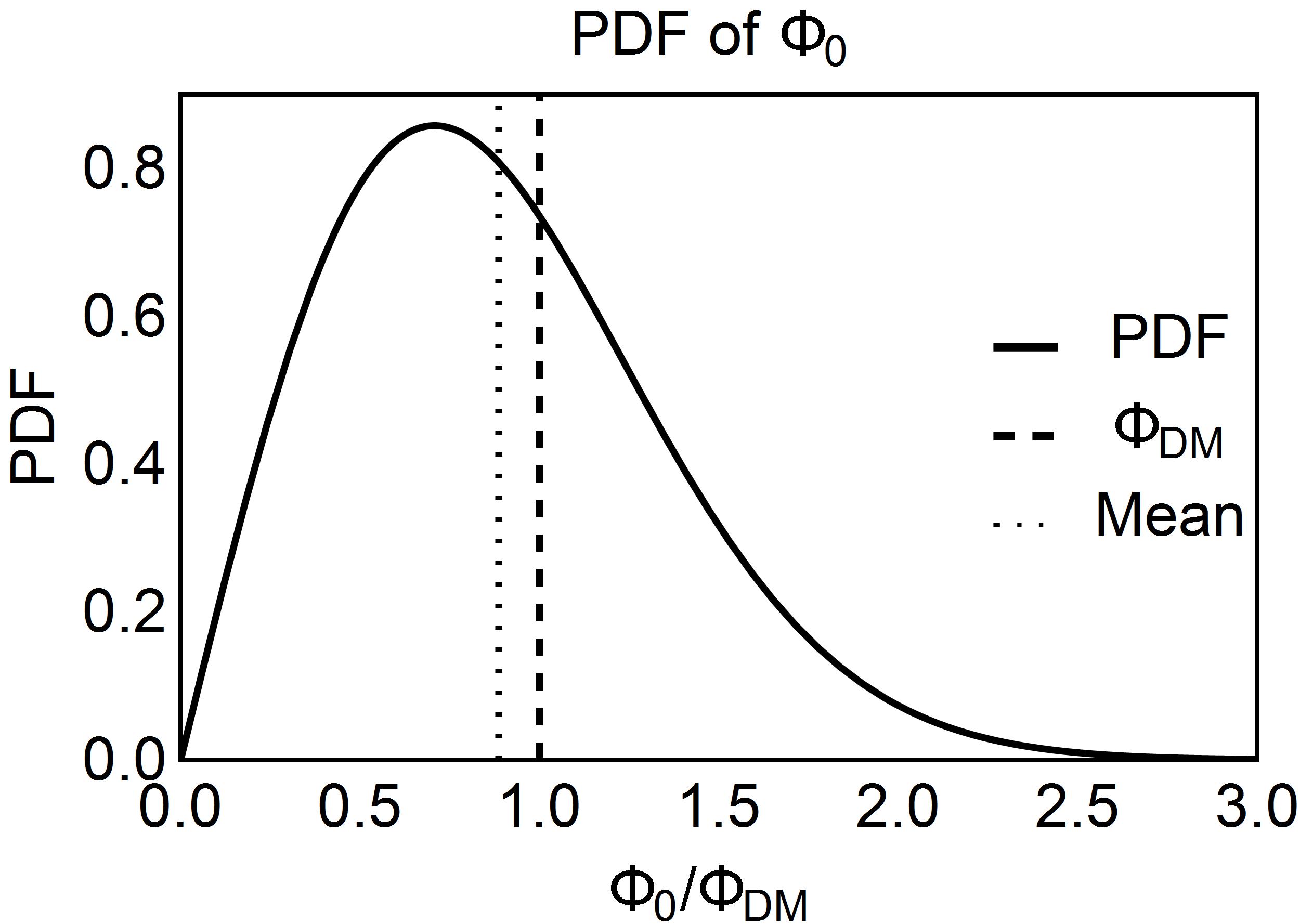}
	\caption{Rayleigh distribution given by Eq.~\eqref{phi0dist}. \label{Fig:Rayleigh} }
\end{figure}

  Note that the average value of this distribution $\langle\Phi_0\rangle=\Phi_{\text{DM}}\sqrt{\pi}/2$ is not simply $\Phi_{\text{DM}}$. Approximately 63\% of given field realizations will have a field amplitude smaller than $\Phi_{{\rm{DM}}}$. The distribution compared to $\Phi_{{\rm{DM}}}$ is shown in Fig.~\ref{Fig:Rayleigh}.
  
\vspace{120pt}

\bibliography{Amplitude_Fluctuations_in_Bosonic_Dark_Matter,library-apd}